# Tunable phononic coupling in excitonic quantum emitters

Adina Ripin[1], Ruoming Peng[2*], Xiaowei Zhang[3], Srivatsa Chakravarthi[1], Minhao He[1], Xiaodong Xu[1,3], Kai-Mei Fu[1,2], Ting Cao[3], and Mo Li[1,2,*]

[1]Department of Physics, University of Washington, Seattle, WA 98195, USA
[2]Department of Electrical and Computer Engineering, University of Washington, Seattle, WA 98195, USA
[3]Department of Material Science and Engineering, University of Washington, Seattle, WA 98195, USA
* pruoming@uw.edu
*moli96@uw.edu

**ABSTRACT**

**Engineering the coupling between fundamental quantum excitations is at the heart of quantum science and technologies. A significant case is the creation of quantum light sources in which coupling between single photons and phonons can be controlled and harnessed to enable quantum information transduction. Here, we report the deterministic creation of quantum emitters featuring highly tunable coupling between excitons and phonons. The quantum emitters are formed in strain-induced quantum dots created in homobilayer semiconductor $WSe_2$. The colocalization of quantum confined interlayer excitons and THz interlayer breathing mode phonons, which directly modulate the exciton energy, leads to a uniquely strong phonon coupling to single-photon emission. The single-photon spectrum of interlayer exciton emission features a single-photon purity >83% and multiple phonon replicas, each heralding the creation of a phonon Fock state in the quantum emitter. Owing to the vertical dipole moment of the interlayer exciton, the phonon-photon interaction is electrically tunable in a wide range, promising to reach the strong coupling regime. Our result demonstrates a new type of solid-state quantum excitonic-optomechanical system at the atomic interface that emits flying photonic qubits coupled with stationary phonons, which could be exploited for quantum transduction and interconnection.**

The quantized vibrational motion in solid-state quantum systems, *i.e.*, phonons, has been exploited as an important modality that interfaces with electrons and photons for quantum information science and applications[1,2]. In bulk diamond, the optical phonon with a frequency of 40 THz has been mapped to Raman scattered photons to realize non-local entanglement at room temperature[3–5]. In microscopic cavity optomechanical systems, phonons of MHz to GHz frequencies have been used to store and transfer quantum states between microwave and optical photons[6,7]. In molecular quantum emitters, coupling with the phonons of the host medium is generally considered detrimental to the quantum properties[8], although the molecule's internal optomechanical degree of freedom has been exploited[9,10]. These archetypical demonstrations utilize either bulk phonon modes involving collective vibration of many atoms or a phonon band including a large number of unresolved modes, resulting in a relatively low phonon-photon scattering probability or coupling rate. To further explore the phonon degree of freedom in the quantum regime, therefore, it is highly desirable to engineer new quantum light sources that afford strong phonon-photon coupling involving a well-defined single phonon mode, preferably with an intermediate high frequency and tunable coupling strength.

Quantum emitters (QEs) in two-dimensional materials[11,12], including transition metal dichalcogenide (TMDs), hexagonal boron nitride (h-BN), and their heterostructures, provide new opportunities for engineering quantum-regime phonon-photon coupling. These atomically-thin materials have bright exciton emissions and are very rich in optically (Raman) active phonon modes[13–16]. Their multilayers further afford intralayer and interlayer phonons with a vast span of frequencies from optical to acoustic ranges. Moreover, QEs in 2D materials can be deterministically created with several approaches, such as strain engineering and ion implantation[17–19]. These techniques have enabled the site-controlled creation of QEs with a high yield and high single-photon purities. Being hosted in atomically thin materials, 2D QEs are amenable to photonic integration to facilitate photon extraction[20] and Purcell effect enhancement of emission[21], providing a versatile quantum system to explore phonon-photon interactions.

Here we report strong and tunable phonon-photon coupling in strain-engineered 2D QEs that are deterministically created in bilayer $WSe_2$. These 2D QEs emit single photons with high purity and have a high electrical tunability in emission energy. Remarkably, in the single-photon emission spectra, multiple well-resolved phonon replica lines are observed, each of which heralds the creation of a phonon Fock state. The large phonon-photon coupling in this QE system stems from the colocalization and quantum confinement of the interlayer excitons and the breathing mode phonons, which directly modulates the exciton energy. The very high phonon-photon coupling strength is characterized by a large and electrically tunable Huang-Rhys factor reaching the highest value realized in solid-state quantum emitter systems[22–27]. The demonstrated strong and tunable single phonon-photon coupling provides an invaluable resource for engineering quantum light emission systems with an additional internal mechanical degree of freedom for quantum information processing.

The bilayer $WSe_2$ is an ideal system to explore the phonon-photon interaction as it affords many Raman-active phonon modes covering a broad frequency range[28,29]. Particularly, as shown in Fig. 1a, the interlayer breathing mode (BM) phonon couples with the interlayer excitons (IXs) strongly, as the electronic excitation of an IX with a vertical dipole moment is modulated directly by the interlayer vibrational state. To explore this coupling at the single-photon level, we use the strain-engineering approach to create QEs in bilayer $WSe_2$ by transferring them onto patterned nanopillars[17,18,30], as illustrated in Fig. 1a. The nanopillars induce local strain that modulates the bandgap and thus creates spatial confinement of IXs in the bilayer $WSe_2$[31], resulting in quantum dots with single-photon emission[32]. Fig. 1c illustrates the potential traps formed by the localized strain in the bilayer $WSe_2$. The optically excited IXs are funneled into the traps where they are bound with natural defects and recombine to emit single photons. In comparison to excitons in monolayer TMDs that have no electric dipole moment, IXs in bilayer $WSe_2$ have a large out-of-plane electric dipole moment[29,33–35]. Therefore, they couple efficiently to both a perpendicular electric field that generates a Stark shift and the interlayer vibration. Stark shifts of the IX energy by more than 50 meV has been achieved in bilayer $WSe_2$ and other TMDs[36,37]. Multiple QEs can be created at each site in the nano-pillar array (Fig. 1b). While the QEs at different sites have different emission energy due to the variation of the local parameters such as the amount of strain and the energy level of the defects, applying a local electric field to each of them using separate gate electrodes can tune their emission energies to be the same.

Fig. 1d shows the optical image of a fabricated device. A stack of bilayer $WSe_2$ encapsulated with two hBN layers (~30 nm), with top and bottom graphite electrodes, is transferred onto a substrate with an array of nanopillars made of $SiO_2$. The top and bottom graphite electrodes enable electrical tuning of the QEs' energy. Fig. 1e shows the density functional theory calculated band structure of bilayer $WSe_2$ in its pristine state and strained state with 1% tensile strain. In pristine bilayer WSe$_2$, indirect bandgap transition Q-K can occur, assisted by various single-phonon or two-phonon processes[29]. When under a sufficient amount of strain, the conduction band minimum is shifted from the Q point to the K point, and the valance band Γ point is shifted up such that direct K-K and indirect K-Γ transitions can become favorable in energy, enabling strong exciton coupling to zero-momentum phonons [29,38–41]. Therefore, depending on the level of local strain, IX species corresponding to either Q-K, K-Γ, or K-K transitions can dominate the photon emission. These IX species have different dipole moments[35], but because their energy is susceptible to the interlayer separation, they all couple to BM phonon strongly.

**Tunable quantum emitters**

We fabricated multiple samples. In each sample, we can find multiple QEs at different nanopillars in the array. All the measurements were performed in a cryostat at a temperature of 10 K. Fig. 2a shows the photoluminescence (PL) spectrum of a QE (QE1) as a function of top gate voltage (bottom gate grounded). The excitation laser is 632.8 nm with fixed power of 15 μW. QE1 remains dark until the gate voltage reaches a negative value of ~-3.5 V, after which a bright, sharp PL peak is observed at 1555 meV with a linewidth of 1.1 meV (full-width at half-maximum (FWHM)). This turn-on behavior may be attributed to the IXs being bound to a donor (or acceptor) type defect that electron (or hole) doping is needed for IX to bound to the defect and form a QE. Both donor and acceptor type defects in WSe$_2$ has been reported previously[42]. Among nineteen IX QEs in five devices (Supplementary Figure 1) we have measured, sixteen of them turn on at negative gate voltage and three of them turn on at positive gate voltage. The gate voltage induces doping because of the top/bottom structure asymmetry of the devices, including the nanopillars at the bottom, the thicker bottom hBN that prevents piercing by the nanopillars. There is also possible non-negligible current leakage through the layers. After QE1 is turned on, its emission can be red-shifted by more than 6.4 meV (~3.3 nm in wavelength) with the application of gate voltage up to -7 V. From the dipole model of field tunability: $\Delta U = p \cdot E$, we calculate QE1's IX dipole moment $p$ to be 0.341 $e$·nm, consistent with previous experimental results for K-Γ IX [29,33,34]. Among nineteen IX QEs we have measured (See Supplementary Table 1), we observed three ranges of dipole moments around 0.32, 0.46, and 0.63 $e$·nm. Respectively, they correspond to the theoretical values of KΓ, QK, and KK IXs. As shown by the DFT-calculated band structure under strain (Fig. 1d) and previously reported experimental results, which type of IX dominates the QE depends on the local strain level[29,35,38,43–46]. Because the local strain level can vary a lot at each nanopillar and the formation of QEs is by the accidental occurrence of a proper defect, the strain levels at different QEs can be very different, thereby leading to different types of IX. Even higher tunability can be achieved with QEs in heterobilayers, such as moiré excitons[47], which have an even larger dipole moment. Fig. 2c shows the PL spectra of another QE (QE2), which behaves differently from IX QEs. QE2 is bright at zero gate voltage with an emission energy of 1550 meV, which remains unchanged with varying gate voltage until 3.0 V when the emission is turned off. This type of QE can be attributed to defect-bound intralayer excitons, which do not couple to the out-of-plane electric field and are rarer (only three out of 22 QEs we measured). However, a sufficiently high electric field can cause electron or hole tunneling to another layer of $WSe_2$, consequently turning off the intralayer exciton emission.

The two types of QEs' different responses to electrical modulation allow us to tune them to the same energy, as summarized in Fig. 2e. In Fig. 2b and d, when gate voltages of -5.0 V and 2.0 V are applied to QE1 and QE2, respectively, both QEs emit photons at 1550 meV with similar linewidths. Fig. 2f shows the second-order photon correlation $g^{(2)}(\tau)$ measured from a device (QE13 in Supplementary Information) behaving similarly to QE1. A filter with ~5.0 nm (9.7 meV) bandwidth was used to select the measurement range. The result shows clear antibunching with $g^{(2)}(0)=0.169\pm0.005$, indicating single-photon purity of 83%. From fitting the autocorrelation data, we estimate the QE lifetime to be 2.0±0.25 ns. We have measured similar antibunching results from many QEs with linewidths in the range of 1 to 3 meV across different devices, which are included in the Supplementary Information. The demonstrated wide electrical tuning range makes these 2D QEs promising for achieving scalable arrays of indistinguishable single-photon sources. In areas without nanopillars, we measure IX emission with a much broader linewidth > 5 meV and without antibunching, which is consistent with previous reports[29,33,35,48]. Also, we observed no pronounced emission in a wide energy range below the bandgap (Supplementary Figure 13), emission due to in-gap defect states can be ruled out. Therefore, we conclude that the combination of defect and strain engineering is necessary to create the QEs[30]. There are several possible types of defects in $WSe_2$, including Se vacancy, W anti-site, O-passivated Se vacancies, and O interstitials[32,49–52]. Although which type of defect is responsible for forming the QE is unclear, requiring microscopic study to reveal, the Se vacancy has a higher density than other types so is more likely to occur at the nanopillars.

**Single exciton-phonon coupling**

A very notable feature in the PL spectra of QE1 (Fig. 2a) is multiple emission lines on both sides of the main peak. These emission lines are turned on/off by the gate voltage along with the main peak and modulated at the same rate, suggesting their correlation. We observed similar features in many devices. Figs. 3a and b show the PL spectra of QE1 and another QE (QE3), measured with gate voltages of -6.4 V and -5.0 V, respectively. Five emission lines can be observed with spacings in the range of 3.0-3.7 meV for QE1 (Fig. 3a), and 3.7-5.1 meV for QE3 (Fig. 3b). We then measured the polarization of the photons from each emission line. As shown in Figs. 3c and d, photons from each QE's emission line are linearly polarized with the same orientation. Figs. 3e and f show that the energies of these lines are tuned by the gate voltage synchronously at the same rate, with their spacings unchanged. All these features allow us to conclude that these emission lines originate from the same QE, rather than other emitters nearby.

The observation of multiple well-resolved emission lines can be explained as phonon replicas due to the coupling between colocalized single IX and a single phonon mode in the QE. Their coupling can be understood with the Franck-Condon principle, as illustrated in Fig. 4a[26,27,53–55]. The ground state of the QE and its excited state when an IX is generated can be modeled with two potential energy surfaces (PES). Each PES is populated with phonon states that are not evenly spaced because of the anharmonicity of the PES. Under the linear coupling approximation, the phonon-exciton coupling is represented by a shifted equilibrium position of the excited-state PES relative to the ground-state. At low temperatures, the QE at the ground state PES has nearly zero phonon occupancy. When an exciton is created by the pump laser, the QE is excited into higher-energy PES and quickly relaxes to its zero phonon level. Upon exciton recombination, the QE emits a single photon and relaxes to its ground-state PES but at an elevated phonon state (Fig. 4a). As a result, the energy of the emitted photon is Stokes-shifted from the zero-phonon line (ZPL), forming phonon replicas spaced by the phonon energy. In Fig. 4a, the emission lines are labeled with the corresponding phonon number state $|n\rangle$ in the ground state PES. According to the Huang-Rhys theory for discrete phonon lines, the intensity of the $n^{\text{th}}$ phonon line is proportional the overlap integral between the initial and final phonon states, i.e., $|\langle 0|n\rangle|^2 = e^{-S}S^n/n!$, where $S$ is the dimensionless Huang-Rhys factor measuring the strength of the exciton-phonon coupling[56]. Therefore, the phonon line intensities have a Poisson distribution with an expectation value of $S$, as illustrated in Fig. 4a. In many other solid-state QEs, such as the color centers in the diamond, the coupling of defect emitters to bulk phonons produces phonon sidebands that are not well resolved due to the continuous phonon DOS in energy. In contrast, in Figs. 3a and b, we observe clearly resolved phonon lines generated by a single localized phonon mode that couples with a single interlayer exciton in the QE.

The energy spacing of the phonon lines in the range of 3.0-5.0 meV matches the energy of the BM phonons in bilayer $WSe_2$. In pristine bilayer $WSe_2$, using time-resolved transmission measurement, the frequency and coherence time of the BM phonon are measured to be 0.8 THz (or 3.4 meV) and 3.5 ps, respectively[54]. However, the interaction between the $WSe_2$ and the encapsulating h-BN can cause mode hybridization so as to modulate BM phonon's frequency[57]. Additionally, the strong local strain gradient and the quantum confinement also affect the BM phonon frequency. As a result, we observe a spread of phonon line spacing in energy in the many devices we measured (See supplementary materials for a comprehensive dataset).

In Fig. 4b, we measured the PL spectra of QE1 with increasingly negative gate voltage and a fixed pump power of 15 µW. In addition to a consistent red-shift of the phonon lines, their intensity distributions have a significant change, indicating the modulation of the phonon coupling strength $S$. We extract the $S$ factors by fitting the peak intensities with the Poisson distribution[27]. Because the PES of the QEs is more complicated than the simple harmonic oscillator model assumed in the Huang-Rhys theory, the fitting has a large error, which outweighs the signal's noise, so it can only estimate the $S$ factors. More details of the fitting method can be found in the Supplementary Information. Despite the relatively large fitting errors, the result of $S$ versus gate voltage in Fig. 3c clearly shows that $S$ initially increases with the gate voltage, reaching a maximal value of 6.3 at ~-5V, and saturates afterward. Although the coupling strength $g_0$ is proportional to the effective field across the materials, the dependence of S on the field is indirect and cannot be ascribed to a simple linear dependence. Especially, when the phonon coupling is large, the dependence of $S$ on the coupling strength and electric field will involve the detailed change of the potential energy surface (PES) in response to the electric field. The saturation is possibly due to gate leakage and charge screening, which prevent further increase of the electric field, as also observable in the emission energy tuning in Fig. 3e. For comparison, we also measured $S$ with increasing pump power and a fixed gate voltage of -6.4 V. $S$ for the most part is unaffected by the pump power. Our results show that the coupling between the colocalized quantum phonon mode and single exciton can be precisely tuned with an electric field to a large value.

The colocalization of the IX and the BM phonon in the strain-engineered QE results in a very large coupling between them. The single exciton-phonon coupling can be described with the spin-boson interaction Hamiltonian: $H_{int,v} = g_0\, \sigma_z(b^\dagger + b)$, where $b\,(b^\dagger)$ is the phonon annihilation (creation) operator on phonon number states $|n\rangle$, $\sigma_z$ is the Pauli matrix on the basis of the ground ($|g\rangle$) and excited ($|e\rangle$) states of the exciton, $g_0$ is the single exciton-phonon coupling rate[1,58]. The BM phonon directly modulates the dipole moment of the IX by adjusting the interlayer distance. Under an electric field $E$, the single-phonon-IX coupling rate[1] can be calculated from the zero-point amplitude ($\bar{x}_{zpf}$) of the Γ-point BM: $g_0 = g^0 + 2eE\bar{x}_{zpf}$, where $\bar{x}_{zpf}$ is calculated with first-principle theory to be ~0.026 Å. $g^0$ is zero-field coupling rate, which is calculated to be <0.08 meV, one order of magnitude smaller than the second term when the field is strong. Thus, assuming that the applied field can reach a reasonably high value of 0.25 V/nm in our device[59], $g_0$ can be increased to 1.3 meV. The system thus can reach the strong-coupling regime, satisfying $2g_0 > (\gamma_e, \gamma_p)$, where $\gamma_e$~1.17 meV is the exciton linewidth (Fig. 2b) and $\gamma_p$<1.18 meV is the BM phonon decoherence rate given its coherence time of 3.5 ps measured at room temperature[36,37,59]. At cryogenic temperatures and with better sample quality, we expect $\gamma_e$ and $\gamma_p$ can be further reduced.

**Conclusion**

We have created highly tunable QEs in WSe$_2$ bilayers with strain engineering methods. An immediate next step is to integrate these emitters with integrated photonic platforms such as waveguides and photonic crystal cavities to achieve super-radiant emission, Purcell enhancement, and cavity QED. With twisted bilayer WSe$_2$ rather than a natural bilayer, the IX energy can be tuned electrically while still maintaining the characteristic valley states similar to those reported in moiré systems[36,37,60]. Furthermore, the demonstrated tunable phonon-photon coupling strength in these 2D QEs provides a new way to explore the quantum optomechanical effects. In many ways, the 2D QEs are analogous to cavity optomechanical systems, but at the atomic interface and with a phonon frequency of 0.8-1.0 THz, which is already at the ground state at the measurement temperature of 10 K. It will be possible to use a THz source to resonantly excite the phonon and prepare phonon Fock states in such a solid-state system at a temperature higher than ever before.

The archetypical excitonic-optomechanical system demonstrated here thus has the potential to be a new quantum light resource that is entangled with single phonons[1,2,60,61] for use in quantum information processing, storage, and communication[1,2,61,62].

**Methods**

*Device Fabrication:* A 17 by 17 array of nanopillars, each with a diameter of 140 nm, a height of 200 nm and spaced with a 2.5 µm pitch, were patterned on 300 nm $SiO_2$ deposited by e-beam evaporation using electron beam lithography. The 2D material heterostructure consists of bilayer $WSe_2$ encapsulated by hBN of thickness around 30nm with few-layer graphite as top and bottom gate electrodes. In some devices, an additional graphite layer is transferred in contact with the $WSe_2$ to control the doping of the material. The heterostructure is stacked and transferred onto the nanopillar array using the conventional dry transfer method using PC/PDMS (Sylgard 184) stamps. After the device has been transferred, contact electrodes of Ti/Au to the graphite were patterned with photolithography using a direct laser writer and deposition using electron beam evaporation. An optical image of a device is shown in Fig. 1d, where the strain induced by the nanopillars is visible.

*Photoluminescence measurement method*: The sample was wire bonded to a sample holder and loaded into a cryostat (Montana Cyrostation S-100) to be cooled to 10K for measurement. A leakage test was done to ensure that the electrodes are not shorted together, and then the photoluminescence spectra of strained heterostructure can be measured at different bias voltages. A continuous-wave He-Ne laser (632.8nm) was used to excite the excitons. The laser beam was focused on the sample through an optical window in the cryostat with a 50× objective lens (NA = 0.42) to achieve a diffraction-limited spot size of about 1 µm in diameter. The interlayer exciton emission was passed through a 633 nm notch filter to remove the reflected signal of the laser before being acquired with a spectrometer (Princeton Instruments, IsoPlane 320).

*Measurement of $g^2(0)$:* Using the same set-up as above, after filtering the laser light with the 633-notch filter, the emission signal was then filtered with a tunable DBR filter with 4 nm bandwidth (810 nm at an incident angle of 0°, and 800 nm at 19°) to select a single emitter line. The signal was then split with a beam splitter and sent to two single-photon detectors (Excelitas, SPCM-AQRH-16) which fed into a time-to-digital converter (ID Quantique) for the time-tagged time-resolved (TTTR) data acquisition.

*First-principles calculations*: Density functional theory calculations were performed using the Quantum Espresso[63] package. The Perdew-Burke-Ernzerhof functional[64] and optimized norm-conserving Vanderbilt pseudopotential (ONCVPSP)[65] were used. The dispersion correction with the D2 form was used to consider the van der Waals interaction[66]. The optimized lattice constant and interlayer distance are 3.334 Å and 6.417 Å, respectively. The phonon dispersion and vibration mode were calculated based on the density-functional perturbation theory. The zero-point amplitude for the breathing mode was calculated as the average of the mean square displacement at the zero temperature, i.e. $\bar{x}_{zpf} = \frac{1}{6}\sum_i \sqrt{\frac{\hbar}{2M_i\omega}}|\xi_i|$, where $\xi_i$ is the eigenvector, $M_i$ is the mass of $i$th atom in the unit cell, and $\omega$ is the phonon frequency.

**Data availability statement**

The data that support the findings of this study are available from the corresponding author upon reasonable request.

**FIGURES**

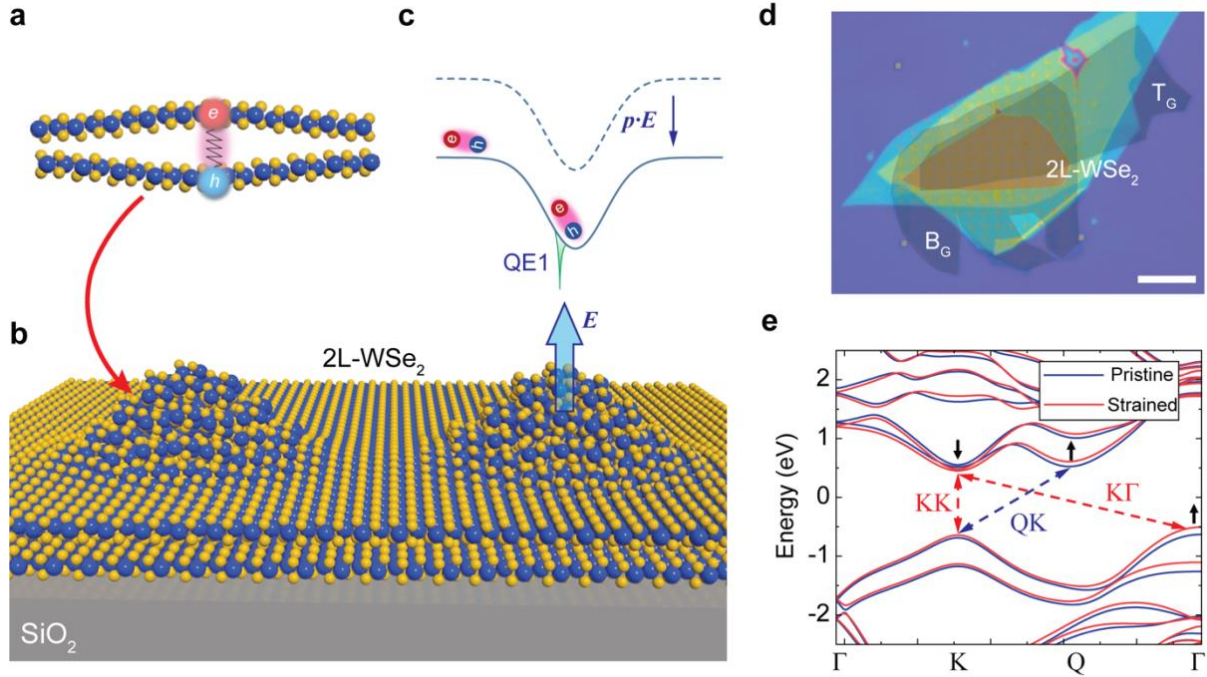

**Figure 1 Strain-engineered 2D quantum emitters (QEs). a.** Schematic of the interlayer breathing mode (BM) phonon which couples strongly with interlayer excitons in bilayer $WSe_2$ because the excitation of the interlayer excitons with vertical dipole moments is directly modulated by the vibrational mode. **b.** Illustration of bilayer $WSe_2$ transferred onto silicon dioxide nanopillars, forming quantum dots with local strain modulation, which host the QEs. **c.** Strain-induced potential traps with natural defects (green dips), where excitons are funneled into and localized. The QE energy can be efficiently tuned by electric fields through the Stark effect. **d.** Optical microscope image of a representative device. The bilayer $WSe_2$, encapsulated by hBN and gated with top and bottom graphite layers ($T_G$, $B_G$), is transferred onto an array of $SiO_2$ nanopillars. Metal electrodes are deposited later to contact the graphite layers. Scale bar: 10 μm. **e.** The DFT calculated band structure of the bilayer $WSe_2$ in pristine (blue) and under 1% tensile strain (red) conditions. Without strain, phonon-assisted indirect bandgap transition Q-K can occur. A sufficient amount of strain can shift the conduction band minimum from the Q point to the K point and shift the valance band Γ point up such that direct K-K and indirect K-Γ transitions can become favorable.

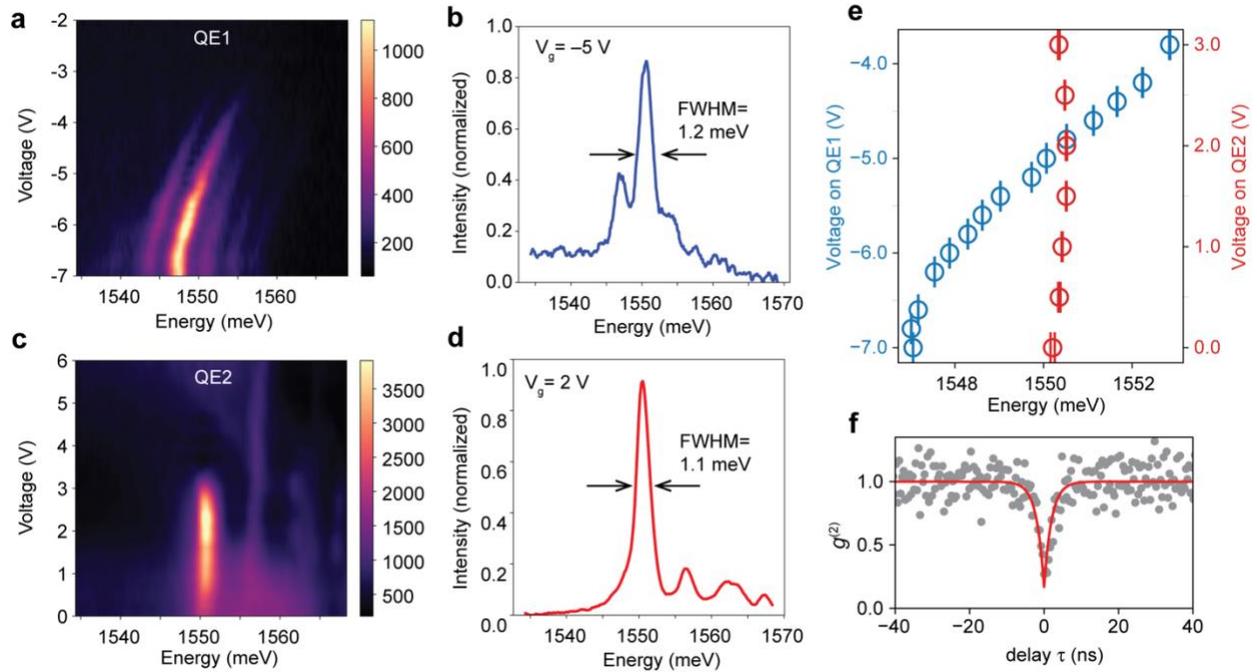

**Figure 2: Tunable 2D QEs. a.** Photoluminescence (PL) spectra of QE1 as a function of the gate voltage. The emitter turns on at around -4V and is accompanied by sidebands that are turned on at the same voltage. **b.** Line-cut of the PL in panel **a** at -5V, showing a full-width half-maximum (FWHM) linewidth of 1.2 meV. **c.** PL spectra of QE2 as a function of the gate voltage. The emitter is on at zero gate bias, suggesting it has a different origin than QE1. At around 3.5V, the emitter is turned off. **d.** Line-cut of the PL in panel **c** at 2V. The linewidth is 1.1 meV. **e.** Modulation of emitter energy of QE1 and 2 with gate voltage. The two QEs can be tuned to have nearly identical energy and linewidth. **f.** Second-order photon correlation $g^{(2)}(\tau)$ measured from a representative QE, showing antibunching with $g^{(2)}(0)=0.169\pm0.005$, indicating single-photon purity of 83%.

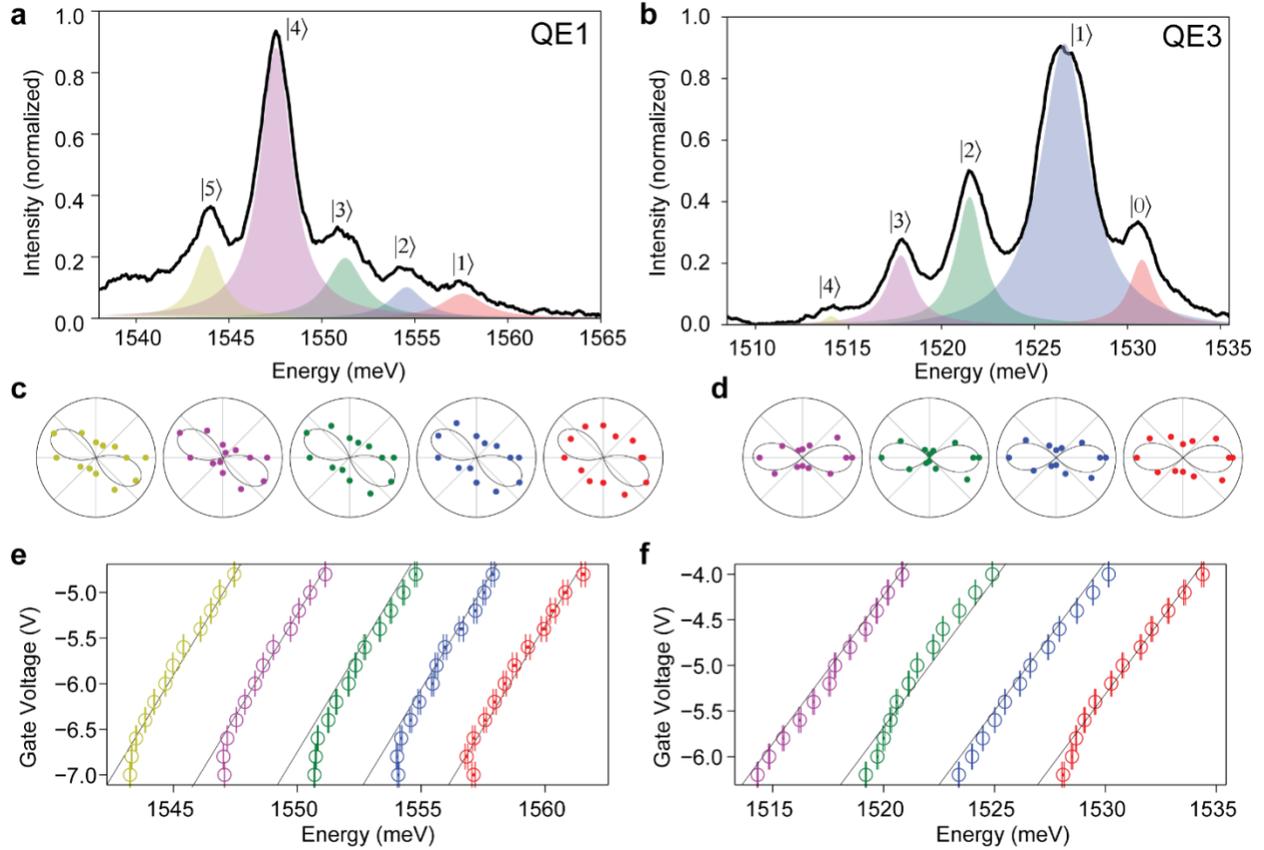

**Fig. 3: Single phonon emission lines. a.** PL spectrum of QE1 at -6.4V. Five emission lines can be observed and are fitted with Lorentzian functions (shaded area). **b.** PL spectrum of QE3 at -5.0 V. Four emission lines can be observed and are fitted with the Lorentzian functions (shaded area). **c, d.** The photons from the sidebands of QE1 (**c**) and QE3 (**d**), respectively, are polarized with the same orientation. The thin black lines are guides for linear polarization. **e, f.** The sideband energy of QE1 (**e**) and QE3 (**f**), respectively, are tuned by the gate voltage synchronously at the same rate with an unchanged spacing in energy.

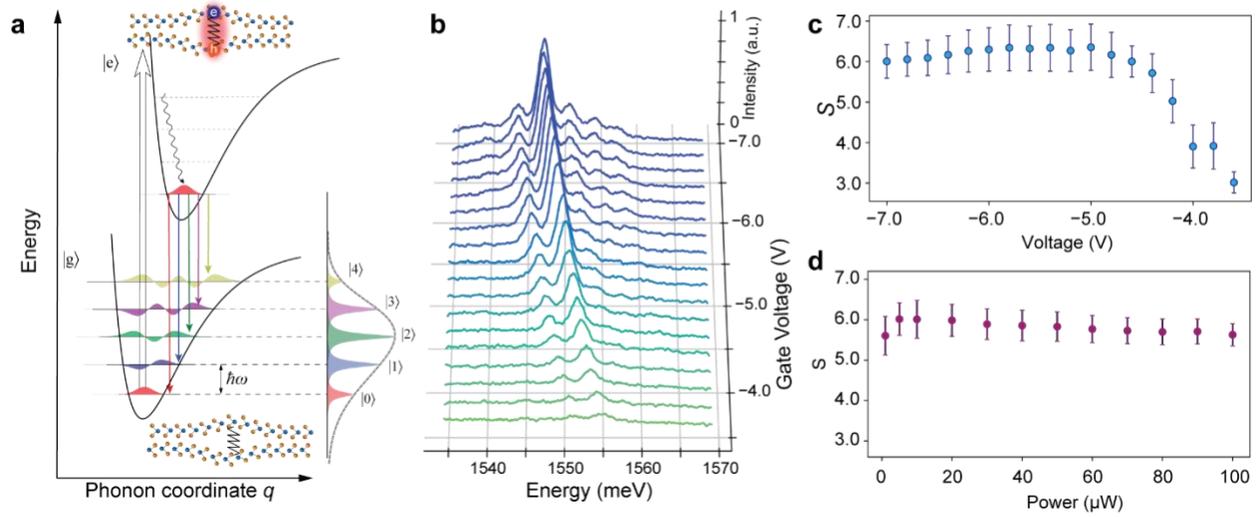

**Figure 4: Strong and tunable single phonon-exciton coupling. a.** A diagram of the Franck-Condon model of exciton-phonon coupling in the 2D QE. The lower parabolic potential represents the ground-state PES of the QE. Multiple phonon states with an energy spacing of 3.4 meV are shown, with their wavefunctions illustrated. The upper potential represents the excited-state PES of the QE with optically pumped IX. The phonon mode that strongly couples with the IX is the breathing mode (BM), in which the layers of $WSe_2$ move in opposite directions, modulating the interlayer spacing and IX energy. Based on the Huang-Rhys theory, the phonon line intensity should have a Poisson distribution with an average phonon number at *S*, which is the Huang-Rhys factor. The anharmonic effect can result in unequal energy spacing for the adjacent phonon lines. **b.** PL spectra of QE1 at different gate voltages. The intensity distribution of the phonon lines is strongly modulated by the gate voltage. We extract the value of *S* by fitting the intensity with the Poisson distribution. **c**. The Huang-Rhys factor (*S*) as a function of the gate voltage. S initially increases with the voltage, in agreement with theory, and levels off after -5.0 V. **d.** The Huang-Rhys factor (*S*) as a function of pump power.